\documentclass[aps,prc,preprint,nofootinbib,superscriptaddress]{revtex4-2}

\usepackage[T1]{fontenc}
\usepackage[utf8]{inputenc}
\usepackage{amsmath,amssymb,amsfonts,mathtools,bm}
\usepackage{graphicx}
\usepackage{xcolor}
\usepackage[colorlinks=true,linkcolor=blue,citecolor=blue,urlcolor=blue]{hyperref}
\usepackage[nameinlink,capitalise]{cleveref}
\usepackage{physics}

\newcommand{\ii}{\mathrm{i}}
\newcommand{\ee}{\mathrm{e}}

\newcommand{\TRI}{time-reversal invariance}

\newcommand{\pw}{\mathrm{pw}}
\newcommand{\tot}{\mathrm{tot}}
\newcommand{\el}{\mathrm{el}}
\newcommand{\reac}{\mathrm{reac}}
\newcommand{\eff}{\mathrm{eff}}
\newcommand{\instate}{\mathrm{in}}
\newcommand{\outstate}{\mathrm{out}}
\newcommand{\zhat}{\hat{\mathbf z}}
\newcommand{\khat}{\hat{\mathbf k}}
\newcommand{\avg}[1]{\left\langle #1 \right\rangle}

\begin{document}

\title{Generalized Optical Theorem for Structured Neutron Beams\\
and Consequences for Forward-Transmission Null Tests of Time-Reversal Invariance}

\author{S.~Samiei}
\affiliation{Department of Physics and Center for Exploration of Energy and Matter, Indiana University, Bloomington, Indiana 47408, USA}

\date{\today}

\begin{abstract}

The simple form of the optical theorem of scattering theory, $\sigma_{\tot}^{(\pw)} = \frac{4\pi}{k}\Im f(0)$, is valid for an incident plane wave or for a wave packet whose Fourier components possess azimuthal symmetry about the incident wave vector $\vec{k}$. Previous work has shown that this expression can break down for structured beams of light which possess orbital angular momentum (OAM), despite the fact that there is clearly no violation of unitarity, and the relevant modifications have been worked out for the case of massless photons. We present a form of the optical theorem involving neutron OAM states for the case of the scattering of massive nonrelativistic particles. We apply this form to Ryndin's theorem on the application of time reversal symmetry to forward scattering, indicate how the statement of the null condition for T violation in forward scattering is modified, and show that this effect is negligible compared with other sources of systematic error in neutron optics transmission experiments. 
\end{abstract}

\maketitle

\section{Introduction}

The optical theorem is among the most familiar identities in scattering theory. In the usual nonrelativistic normalization it reads
\begin{equation}
\sigma_{\tot}^{(\pw)}=\frac{4\pi}{k}\,\Im f(0),
\label{eq:pw_ot_intro}
\end{equation}
where $\sigma_{\tot}$ is the total cross section, $k$ is the incident wave number, and $f(0)$ is the zero-angle scattering amplitude. The theorem is fundamentally a statement of unitarity, and the plane-wave derivation leading to the expression~\cref{eq:pw_ot_intro} is largely unchanged for azimuthally-symmetric nonrelativistic wave packets that do not suffer wave packet distortion in the interaction~\cite{GoldbergerWatson1964,Taylor1972,Newton1982}. In this paper we explore what changes if the prepared incident state is \lq\lq structured\rq\rq. Over the last couple of decades it has become possible to experimentally engineer a wide variety of more complicated types of wave packet superpositions for different particles~\cite{allen1992orbital, mair2001entanglement, Andersen2006, wang2012terabit, karlovets2015, rubinsztein2016roadmap, ritsch2017orbital, shen2019optical, Ivanov2022, bliokh2023roadmap}. In this paper we will focus on a particular form of structured wave packets referred to in the scientific literature as orbital angular momentum (OAM) states for neutrons~\cite{Clark2015}, which we describe in more detail below. Since these wave packet superpositions need not obey azimuthal symmetry, the question arises of how the unitarity condition for scattering should be formulated in this case. 

This question can matter for polarized neutron optics involving parity and time reversal violation. It is easy to see the source of the issue. Some neutron-nucleus interactions can generate neutron OAM in the final state after transmission through the medium, even if it is absent in the initial state. Consider for example a parity-odd term in the neutron forward optical limit of the form $\vec{\sigma} \cdot \vec{k}$ where $\vec{\sigma}$ is the neutron spin~\cite{Michel1964}. Such a term can give rise to parity-odd neutron optical rotary power, which would rotate the plane of polarization of a transversely-polarized neutron about the neutron momentum even for neutron scattering from an unpolarized ensemble of spin $0$ nuclei. Simply by the conservation of angular momentum, it is clear that the outgoing neutron wave must develop a nonzero orbital angular momentum, and the final state in this case is a particular type of OAM state. Other neutron-nucleus interactions~\cite{Afanasev2019, Afanasev2021, Jach2022}, electromagnetic neutron interactions~\cite{Geerits2021, Geerits2025}, and neutron interactions with condensed matter~\cite{sherwin2022scattering} can also generate neutron OAM, and methods to create and analyze neutron OAM states are under development~\cite{nsofini2016spin, sarenac2018methods, Cappelletti2018,  sarenac2019generation, sarenac2022experimental,  geerits2023phase, le2023spin, sarenac2024small, sarenac2025generation, mckay2025topological}. Recent detailed discussion of structured neutron states can be found in the literature~\cite{Geerits2025, lailey2025multimode}. 

We are especially interested in how this phenomenon may influence the analysis of polarized neutron optics experiments on time reversal invariance. In the neutron optics literature, one writes the T-odd and P-odd difference of total cross sections as
\begin{equation}
\Delta\sigma_{TP}=\frac{4\pi}{k}\,\Im\!\bigl(f_{\uparrow}-f_{\downarrow}\bigr),
\label{eq:delta_sigma_TP_intro}
\end{equation}
where $f_{\uparrow,\downarrow}$ are the zero-angle forward scattering amplitudes for neutron spin parallel and antiparallel to the axis $\khat\times\mathbf I$ defined by the neutron momentum $\vec{k}$ and target polarization $\vec{I}$. Long ago Ryndin et al.~\cite{BilenkiiLapidusRyndin1965,BilenkiiLapidusRyndin1969,BowmanGudkov2014} proved a key null theorem in scattering theory for T violation in the forward limit which seems not to be widely appreciated: in forward elastic transmission, $\Delta\sigma_{TP}=0$ if $f_{\uparrow}$ and $f_{\downarrow}$ transform into each other under time reversal. Later Bowman and Gudkov~\cite{BowmanGudkov2014} applied this theorem to an explicit experimental example and showed that this theorem could be realized experimentally.  The scientific significance of this null result is that it enables sensitive searches for the violation of time reversal symmetry in neutron-nucleus reactions. Searches for new sources of time reversal violation are intensively pursued for the insight they may provide on the origin of the baryon asymmetry of the universe and on physics beyond the standard model of particles and interactions~\cite{Flambaum1999, pospelov2005electric, Engel2013, Roberts2015, abel2020measurement, deVries:2020iea, Engel2025}, and polarized neutron transmission measurements in polarized and aligned nuclear targets can be conducted with high sensitivity to new T violation sources~\cite{Gudkov:1990tb, flambaum2022parity}

It is important to emphasize that the underlying time-reversal statement is older and more general than any particular plane-wave representation. In the earlier literature, time-reversal invariance was formulated first as a constraint on the scattering operator, i.e.\ at the $S$-matrix level, and hence as a relation between direct and inverse processes; only afterward was it specialized to scattering amplitudes and polarization observables~\cite{BilenkiiLapidusRyndin1965,BilenkiiLapidusRyndin1969}. The present work therefore does not propose a new symmetry principle. Rather, it identifies the correct realization of that operator-level statement for structured massive-particle beams, where the relevant forward quantity is the amplitude for the neutron to remain in the same prepared incident mode, and where the appropriate time-reversed comparison must be taken in the full mode space. In this sense our result is a reformulation, for structured neutron beams, of the null-test logic that appears in its conventional forward-transmission form in the standard neutron-optics setting~\cite{BowmanGudkov2014}.
 
Since the derivation of this null-test logic is built on the standard forward optical theorem, the presence of structured beams in the initial or final state of the scattering system raises a conceptual question: what becomes of \cref{eq:pw_ot_intro} and \cref{eq:delta_sigma_TP_intro} when the incident neutron is not a plane wave? We already know that modifications are required in general for structured light. Over the last three decades structured waves referred to as orbital angular momentum (OAM) states have drawn more attention. In light optics this subject began with the recognition that Laguerre-Gaussian modes can carry well-defined orbital angular momentum~\cite{allen1992orbital}, and has since developed into a broad field of structured light and vortex beam studies~\cite{mair2001entanglement, karlovets2015, rubinsztein2016roadmap, shen2019optical, Ivanov2022, bliokh2023roadmap}. It was soon shown experimentally by Krasavin \emph{et al.}\ that the conventional forward-amplitude form of the optical theorem can fail for radially and azimuthally polarized beams of this type. Of course unitarity remains intact, and a properly-generalized optical theorem still predicts the extinction cross section correctly~\cite{Krasavin2018}. Still this work showed clearly that the textbook expressions need not be correct for a structured beam.

Krasavin \emph{et al.} studied scattering from subwavelength particles illuminated by focused radially polarized beams and showed, experimentally in both the optical and microwave domains, that the textbook optical theorem can fail completely: the scatterer exhibits substantial extinction even when the usual forward scattered field vanishes by symmetry~\cite{Krasavin2018}. In their case the problem is tied to the vector nature of the electromagnetic field, because strong longitudinal field components on the beam axis excite a dipole along the propagation direction that radiates mainly away from the forward axis, so the standard 
formula misses this contribution to the extinction. They resolved this issue by replacing the usual textbook expression for the optical theorem with a generalized optical-theorem formulation written for arbitrary vector beams. They proposed a plane-wave-decomposition form of the theorem that correctly reproduces the extinction cross section. For tightly focused beams they also discuss extinction in terms of the power removed from the beam. 


Similar issues should also be present for massive particles. Unlike photons, whose helicity is constrained by the masslessness of the photon, the spin of a massive particle is not kinematically locked to its momentum direction. A structured massive-particle state may therefore carry independently-specifiable spin and orbital content. Our modification will therefore not be in the form of a transversality correction as for vector light beams as encountered by Krasavin \emph{et al.}\: rather we must identify the correct forward amplitude for a structured beam of a massive spin-$1/2$ particle.

For the purposes of the present paper we will not need to delve into the full variety of neutron OAM states. For now we just note that that these states are most conveniently expressed in terms of cylindrical waves and possess nontrivial orbital angular momentum quantum numbers $l$ as eigenfunctions of the usual quantum operator $\hat{\mathbf L}=\hat{\mathbf r}\times\hat{\mathbf p}$ and of the free Schrodinger equation which break azimuthal symmetry. Recent work has established the generation, control, and detection of neutron OAM and related structured neutron states~\cite{Clark2015, nsofini2016spin, sarenac2018methods, Cappelletti2018,  sarenac2019generation, Geerits2021, sarenac2022experimental,  geerits2023phase, le2023spin, sarenac2024small, sarenac2025generation, mckay2025topological, Geerits2025}.
Our results will apply to any massive nonrelativistic spin-$1/2$ particle prepared in a structured incoming mode, but the discussion will focus on the neutron case since neutron optics uses the optical theorem explicitly~\cite{Sears1989} and because forward neutron transmission is where the Ryndin null-test logic for time reversal symmetry has especially sharp physical consequences~\cite{BowmanGudkov2014}. 


The purpose of this paper is therefore twofold. We derive the optical theorem directly in Hilbert space, recover the plane-wave result as a corollary, and formulate the correct version of the optical theorem for a structured incoming beam. We then apply this result to forward neutron transmission and show that the Ryndin null theorem for $T$ symmetry survives if the two relevant amplitudes are true time reverses of each other in the full mode space. We then discuss the effect of imperfect mode reversal which can be induced through interactions of the neutron beam with the target and argue that it is negligible.


The manuscript is organized as follows. In \cref{sec:hilbert} we derive the optical theorem as a Hilbert-space identity and isolate the role of the flux convention. In \cref{sec:plane_wave} we recover the standard plane-wave result. In \cref{sec:structured} we formulate the theorem for structured massive-particle beams, with emphasis on neutron OAM states, and clarify the relation between the coherent same-mode theorem and the incoherent bulk-transmission regime relevant to neutron optics. In \cref{sec:spin_operator} we connect the formalism to the usual spin-dependent
forward operator. In \cref{sec:ryndin} we show how the forward-transmission null theorem survives. In \cref{sec:mismatch} we estimate the effect of imperfect mode reversal. The discussion and outlook are given in \cref{sec:discussion}, and the channel-space form appropriate to noncentral spin-dependent interactions is summarized in \cref{app:noncentral}.

\section{Optical theorem as a Hilbert-space identity}
\label{sec:hilbert}

\subsection{General derivation from unitarity}

Let $S$ be the scattering operator on the asymptotic Hilbert space of scattering states, with
\begin{equation}
S=1+\ii T,
\qquad
S^{\dagger}S=1.
\label{eq:S_def}
\end{equation}
Introduce a complete basis of asymptotic out-states,
\begin{equation}
1_{\outstate}=\sum_f \ket{f}_{\outstate}\,{}_{\outstate}\bra{f},
\label{eq:completeness}
\end{equation}
where the sum includes integrals over continuous labels. Unitarity implies
\begin{equation}
(1-\ii T^{\dagger})(1+\ii T)=1
\quad\Longrightarrow\quad
\ii(T-T^{\dagger})=T^{\dagger}T.
\label{eq:unitarity_T}
\end{equation}
Taking the expectation value in an arbitrary normalized incident state $\ket{\psi}_{\instate}$ gives
\begin{equation}
2\,\Im\,\mel{\psi_{\instate}}{T}{\psi_{\instate}}
=\mel{\psi_{\instate}}{T^{\dagger}T}{\psi_{\instate}}.
\label{eq:master_identity_1}
\end{equation}
Inserting \cref{eq:completeness} between $T^{\dagger}$ and $T$ yields
\begin{align}
2\,\Im\,\mel{\psi_{\instate}}{T}{\psi_{\instate}}
&=\sum_f \mel{\psi_{\instate}}{T^{\dagger}}{f_{\outstate}}{}_{\outstate}\mel{f}{T}{\psi_{\instate}}
\notag\\
&=\sum_f \left|{}_{\outstate}\mel{f}{T}{\psi_{\instate}}\right|^2.
\label{eq:hilbert_ot}
\end{align}
Equation~\eqref{eq:hilbert_ot} is the optical theorem in its most general form. It is an inclusive
identity: the right-hand side is a complete sum over all out-states. At this stage there is no
reference to plane waves, a beam axis, or a specific cross-section normalization.

\subsection{Cross sections and the flux convention}

Equation~\eqref{eq:hilbert_ot} is exact, but by itself it is not yet a cross-section formula. To make
contact with measurable transmission one must specify both the state normalization and the incident
flux convention. Operationally, for any prepared incoming state $\ket{\psi_{\instate}}$ traversing a
target of number density $n_T$ with incident flux $\Phi_{\instate}$, we define
\begin{equation}
\sigma_{\tot}[\psi_{\instate}]
\equiv
\frac{\text{rate of removal from the prepared incident mode}}{n_T\,\Phi_{\instate}}.
\label{eq:cross_section_def}
\end{equation}
This is the same operational definition used for both plane waves and structured beams. For the
transmission problems considered below, $\Phi_{\instate}$ should be understood as the longitudinal
incident flux of the prepared beam through the target aperture. What changes for a structured beam is
not the definition of the cross section, but the forward object to which unitarity relates that loss
rate. The conceptual order is therefore
\begin{equation}
\text{unitarity identity}
\quad\Longrightarrow\quad
\text{forward matrix element}
\quad\Longrightarrow\quad
\text{cross section with flux convention}.
\end{equation}
Keeping these steps distinct is essential once the incident state is no longer a single plane wave.

\subsection{Inclusive versus post-selected observables}

The inclusive character of \cref{eq:hilbert_ot} is central. If one post-selects only a subset
$\mathcal P$ of final states, the observed probability is
\begin{equation}
P_{\mathcal P}
=\sum_{f\in\mathcal P}\left|{}_{\outstate}\mel{f}{T}{\psi_{\instate}}\right|^2
=\mel{\psi_{\instate}}{T^{\dagger}\Pi_{\mathcal P}T}{\psi_{\instate}},
\label{eq:exclusive_prob}
\end{equation}
with projector
\begin{equation}
\Pi_{\mathcal P}=\sum_{f\in\mathcal P}\ket{f}_{\outstate}{}_{\outstate}\bra{f}.
\end{equation}
Unless $\Pi_{\mathcal P}$ is the identity on the full out-space, one does not obtain the optical
theorem by replacing the inclusive sum with $P_{\mathcal P}$. This is precisely why the theorem
governs total transmission loss rather than arbitrarily post-selected channels.

\section{Plane-wave optical theorem as a corollary}
\label{sec:plane_wave}

To recover the familiar form, specialize to plane waves with the standard continuum normalization
\begin{equation}
\braket{\mathbf k'}{\mathbf k}=(2\pi)^3\delta^{(3)}(\mathbf k'-\mathbf k)
\label{eq:momentum_norm}
\end{equation}
and define the on-shell $T$ matrix by
\begin{equation}
\mel{\mathbf k_f}{T}{\mathbf k_i}
=(2\pi)\delta(E_f-E_i)\,T(\mathbf k_f,\mathbf k_i).
\label{eq:T_matrix_def}
\end{equation}
For nonrelativistic potential scattering one may define the amplitude through
\begin{equation}
f(\Omega)=-\frac{\mu}{2\pi}T(\mathbf k_f,\mathbf k_i),
\qquad
|\mathbf k_f|=|\mathbf k_i|=k,
\label{eq:f_from_T}
\end{equation}
so that
\begin{equation}
\dv{\sigma}{\Omega}=|f(\Omega)|^2.
\label{eq:differential_cross}
\end{equation}

For a central interaction the amplitude admits the partial-wave expansion
\begin{equation}
f(\Omega)=\sum_{\ell=0}^{\infty}(2\ell+1)f_\ell P_\ell(\cos\theta),
\qquad
f_\ell=\frac{S_\ell-1}{2\ii k}.
\label{eq:partial_wave_amp}
\end{equation}
If only elastic scattering is open, $|S_\ell|=1$ and one recovers
\begin{equation}
\Im f_\ell=k|f_\ell|^2.
\label{eq:elastic_pw_unitarity}
\end{equation}
More generally, with reaction channels open,
\begin{equation}
S_\ell=\eta_\ell\,\ee^{2\ii\delta_\ell},
\qquad 0\le \eta_\ell\le 1,
\end{equation}
which gives
\begin{equation}
\Im f_\ell=k|f_\ell|^2+\frac{1-\eta_\ell^2}{4k}.
\label{eq:inelastic_pw_identity}
\end{equation}
The forward amplitude is
\begin{equation}
f(0)=\sum_{\ell=0}^{\infty}(2\ell+1)f_\ell.
\label{eq:f0_pw}
\end{equation}
Using \cref{eq:inelastic_pw_identity},
\begin{align}
\Im f(0)
&=\sum_\ell(2\ell+1)\Im f_\ell \notag\\
&=\sum_\ell(2\ell+1)\left[k|f_\ell|^2+\frac{1-\eta_\ell^2}{4k}\right].
\label{eq:imf0_sum}
\end{align}
Meanwhile,
\begin{equation}
\sigma_{\el}=\int \dd\Omega\,|f(\Omega)|^2
=4\pi\sum_\ell(2\ell+1)|f_\ell|^2,
\label{eq:sigma_el_pw}
\end{equation}
and
\begin{equation}
\sigma_{\reac}=\frac{\pi}{k^2}\sum_\ell(2\ell+1)(1-\eta_\ell^2).
\label{eq:sigma_reac_pw}
\end{equation}
Therefore
\begin{equation}
\frac{4\pi}{k}\Im f(0)=\sigma_{\el}+\sigma_{\reac}\equiv\sigma_{\tot}^{(\pw)}.
\label{eq:pw_ot_final}
\end{equation}
This is the standard plane-wave optical theorem, now exhibited as a corollary of the Hilbert-space
identity.

\section{Structured beams and the same-mode forward object}
\label{sec:structured}

\subsection{Wave-packet point of view and OAM modes}

A rigorous scattering theory is naturally formulated in terms of normalizable wave packets
\cite{GoldbergerWatson1964,Taylor1972}. Structured neutron beams therefore lie well within the
standard scattering framework: they are simply specially prepared incoming packet states. A realistic
neutron OAM beam is not a plane wave but a packet whose momentum-space distribution is concentrated
on a cone about a mean propagation axis and carries an azimuthal phase.

Taking the beam axis to be $z$, one may write a normalizable OAM packet as
\begin{equation}
\ket{\phi^{(m)}_s}
=\int\frac{\dd^3k}{(2\pi)^3}
\,g_s(k_\perp,k_z)\,\ee^{\ii m\phi_k}\,\ket{\mathbf k,s},
\qquad
\sum_s\int\frac{\dd^3k}{(2\pi)^3}|g_s|^2=1,
\label{eq:oam_packet}
\end{equation}
where $m$ is the OAM winding number about the beam axis and $s$ labels the spin state. The ideal
Bessel-beam limit corresponds formally to a sharply peaked distribution on a cone, but the packet
form \eqref{eq:oam_packet} is both physically realistic and mathematically cleaner.

Since OAM is part of the prepared asymptotic state, the generalized optical theorem should be a statement about that prepared mode. Since the neutron is massive, the spin label and the orbital
label $m$ need not be kinematically tied to one another. One may work in a spin basis, a helicity
basis, or a coupled basis as convenient, but that is a representation choice, not a massless-particle
constraint.

\subsection{Coherent same-mode amplitude}

To isolate the generalized forward amplitude, fix the momentum magnitude $k$ and represent the incident
mode by its angular content,
\begin{equation}
\ket{\psi_{\instate}}=\sum_s\int \dd\Omega\,a_s(\Omega)\,\ket{\khat(\Omega),s}_{\instate},
\qquad
\sum_s\int\dd\Omega\,|a_s(\Omega)|^2=1.
\label{eq:incoming_mode}
\end{equation}
For a central spin-independent interaction the scattering kernel depends only on the angle between
incoming and outgoing directions,
\begin{equation}
K(\Omega,\Omega')\equiv f\!\left(\khat(\Omega)\cdot\khat(\Omega')\right),
\label{eq:kernel_def}
\end{equation}
and the natural same-mode forward amplitude is the quadratic functional
\begin{equation}
F[a]\equiv
\int\dd\Omega\int\dd\Omega'\,a^*(\Omega)K(\Omega,\Omega')a(\Omega').
\label{eq:F_def}
\end{equation}
For a plane wave, $a(\Omega)=\delta(\Omega-\Omega_0)$ and $F[a]\to f(0)$. Thus $F[a]$ is precisely
the amplitude that replaces the usual forward amplitude in the structured-beam case.

\subsection{Diagonal form for a central interaction}

Expand the incoming mode in spherical harmonics,
\begin{equation}
a(\Omega)=\sum_{\ell m}a_{\ell m}Y_{\ell m}(\Omega),
\qquad
\sum_{\ell m}|a_{\ell m}|^2=1.
\label{eq:a_harmonics}
\end{equation}
Using
\begin{equation}
f(\khat\cdot\khat')=
\sum_{\ell=0}^{\infty}(2\ell+1)f_\ell P_\ell(\khat\cdot\khat')
\label{eq:f_legendre}
\end{equation}
and the addition theorem
\begin{equation}
P_\ell(\khat\cdot\khat')=
\frac{4\pi}{2\ell+1}\sum_{m=-\ell}^{\ell}Y_{\ell m}(\Omega)Y^*_{\ell m}(\Omega'),
\label{eq:addition_theorem}
\end{equation}
one finds (see \cref{app:diag})
\begin{equation}
F[a]=4\pi\sum_{\ell m}f_\ell |a_{\ell m}|^2.
\label{eq:F_diag}
\end{equation}
Using \cref{eq:inelastic_pw_identity},
\begin{align}
\Im F[a]
&=4\pi\sum_{\ell m}|a_{\ell m}|^2\Im f_\ell \notag\\
&=4\pi\sum_{\ell m}|a_{\ell m}|^2\left(k|f_\ell|^2+\frac{1-\eta_\ell^2}{4k}\right).
\label{eq:ImF_diag}
\end{align}
Introducing the mode-dependent elastic and reaction cross sections,
\begin{align}
\sigma_{\el}[a]&\equiv 4\pi\int\dd\Omega\,|(Ka)(\Omega)|^2,
\label{eq:sigma_el_a}\\
\sigma_{\reac}[a]&\equiv \frac{4\pi^2}{k^2}\sum_{\ell m}|a_{\ell m}|^2(1-\eta_\ell^2),
\label{eq:sigma_reac_a}
\end{align}
one obtains the structured-beam optical theorem
\begin{equation}
\sigma_{\tot}[a]=\frac{4\pi}{k}\Im F[a].
\label{eq:mode_ot}
\end{equation}
Equation~\eqref{eq:mode_ot} is the coherent same-mode form of the optical theorem. It states that
for a structured incident beam, ``forward'' means return to the same prepared mode.

\subsection{Which regime is relevant for neutron transmission?}

The coherent same-mode theorem is the cleanest formal statement, but neutron transmission through a
macroscopic target is often described more naturally by an effectively incoherent angular weight
$w(\Omega)\ge 0$ with
\begin{equation}
\int\dd\Omega\,w(\Omega)=1.
\end{equation}
Impact-parameter averaging, source incoherence, and finite acceptance can wash out relative phases
between distinct incident directions. In that case the experimentally relevant forward quantity is the
mode-averaged operator
\begin{equation}
f_{\eff}[w]\approx\int\dd\Omega\,w(\Omega)\,f\bigl(\khat(\Omega)\bigr),
\label{eq:feff_weight}
\end{equation}
and the optical theorem takes the practical form
\begin{equation}
\sigma_{\tot}[w]\approx\frac{4\pi}{k}\Im f_{\eff}[w].
\label{eq:feff_ot}
\end{equation}
This is the regime that is most directly relevant to the forward-transmission applications discussed
below. In other words, the paper first derives the fully coherent same-mode theorem and then applies
its incoherent bulk-transmission reduction to the neutron-optics problem.

For spin-dependent and noncentral interactions, the same-mode forward object becomes channel-valued
rather than diagonal in ordinary orbital partial waves. The basic theorem is unchanged, but the
kernel becomes an operator in combined angular-momentum and spin space. Because the T-odd structures
of interest in neutron transmission are of exactly this type, we summarize the appropriate
channel-space form in \cref{app:noncentral}. The central-potential treatment above should therefore be
read as the simplest transparent realization of the theorem, not as a restriction on its validity.

\section{Mode-effective forward operator in neutron spin space}
\label{sec:spin_operator}

For a plane-wave incident neutron moving along $\khat$ and scattering from a polarized target with
spin operator $\mathbf I$, the forward amplitude is a $2\times 2$ operator in neutron spin space. To
linear order in the target spin one may write
\begin{equation}
\begin{aligned}
f(\khat)={}&A+B(\bm\sigma\cdot\mathbf I)
+C(\bm\sigma\cdot\khat)
+D\,\bm\sigma\cdot(\khat\times\mathbf I)
\\
&+E(\khat\cdot\mathbf I)
+F(\khat\cdot\mathbf I)\,\bm\sigma\cdot(\khat\times\mathbf I),
\end{aligned}
\label{eq:forward_operator_plane}
\end{equation}
where $A,\ldots,F$ are functions of energy. The T-odd
correlation of interest in neutron-transmission null tests resides in the term proportional to
$\bm\sigma\cdot(\khat\times\mathbf I)$ and in related operator structures~\cite{BowmanGudkov2014}. Explicit expressions for $A,\ldots,F$ and discussions of their consequences for neutron spin dynamics have been developed for neutron-nucleus resonance reactions~\cite{Gudkov2017, Gudkov2018, Gudkov2020}.

In the bulk-transmission regime of \cref{eq:feff_weight,eq:feff_ot}, the effective forward operator is
\begin{equation}
f_{\eff}[w]=\int\dd\Omega\,w(\Omega)f\bigl(\khat(\Omega)\bigr).
\label{eq:feff_operator_def}
\end{equation}
For the remainder of this subsection we suppress the spectator spin label and focus on the angular
structure of the forward operator. Introduce the beam moments
\begin{equation}
\avg{\khat}\equiv\int\dd\Omega\,w(\Omega)\khat(\Omega),
\qquad
\avg{\khat_i\khat_j}\equiv\int\dd\Omega\,w(\Omega)\khat_i(\Omega)\khat_j(\Omega).
\label{eq:beam_moments}
\end{equation}
Then
\begin{align}
\int\dd\Omega\,w(\Omega)(\khat\cdot\mathbf I)&=\avg{\khat}\cdot\mathbf I,
\label{eq:moment_scalar}\\
\int\dd\Omega\,w(\Omega)\bm\sigma\cdot(\khat\times\mathbf I)
&=\bm\sigma\cdot\bigl(\avg{\khat}\times\mathbf I\bigr),
\label{eq:moment_vector}
\end{align}
and
\begin{equation}
\int\dd\Omega\,w(\Omega)(\khat\cdot\mathbf I)(\khat\times\mathbf I)
=\bigl([\avg{\khat\khat^{\mathsf T}}\,\mathbf I]\times\mathbf I\bigr).
\label{eq:moment_tensor_identity}
\end{equation}
Substituting into \cref{eq:forward_operator_plane} gives
\begin{equation}
\begin{aligned}
f_{\eff}[w]={}&A+E\,\avg{\khat}\cdot\mathbf I
\\
&+\bm\sigma\cdot\Bigl[B\mathbf I+C\avg{\khat}
+D\bigl(\avg{\khat}\times\mathbf I\bigr)
+F\bigl([\avg{\khat\khat^{\mathsf T}}\,\mathbf I]\times\mathbf I\bigr)\Bigr].
\end{aligned}
\label{eq:feff_operator_final}
\end{equation}
This makes the key point transparent: a structured beam does not by itself create new spin
structures in the forward amplitude. Rather, it replaces the explicit beam direction $\khat$ by beam moments. Accordingly, the
quantities extracted experimentally from a structured-beam transmission measurement are, in general,
mode-effective combinations of the underlying plane-wave coefficients $A,\ldots,F$, not the latter
coefficients themselves. The distinction must be kept explicit when comparing structured-beam results
to plane-wave neutron-optics formulae.

\subsection{Axially symmetric cone relevant to Bessel/OAM beams}

For an axially symmetric cone about the $z$ axis with fixed polar angle $\theta_0$ and uniform
azimuth,
\begin{equation}
w(\Omega)=\frac{1}{2\pi}\,\frac{\delta(\theta-\theta_0)}{\sin\theta_0}.
\label{eq:cone_weight}
\end{equation}
The moments are
\begin{equation}
\avg{\khat}=\cos\theta_0\,\zhat,
\label{eq:first_cone_moment}
\end{equation}
and
\begin{equation}
\avg{\khat_i\khat_j}
=\mathrm{diag}\!\left(\frac{\sin^2\theta_0}{2},\frac{\sin^2\theta_0}{2},\cos^2\theta_0\right).
\label{eq:second_cone_moment}
\end{equation}
Hence
\begin{equation}
\avg{\khat}\times\mathbf I=\cos\theta_0\,\zhat\times\mathbf I,
\label{eq:D_cone}
\end{equation}
and, for axial symmetry,
\begin{equation}
[\avg{\khat\khat^{\mathsf T}}\,\mathbf I]\times\mathbf I
=\frac{3\cos^2\theta_0-1}{2}\,I_z\,(\zhat\times\mathbf I).
\label{eq:F_cone}
\end{equation}
The $D$ and $F$ structures therefore align along the same pseudovector $\zhat\times\mathbf I$, but
with different cone-angle prefactors. This is a concrete example of how structured-beam geometry can
rescale or mix sensitivities if the data are analyzed under a naive plane-wave assumption.

\section{Time reversal and survival of the forward-transmission null test}
\label{sec:ryndin}

We now address the main symmetry question. Does the generalized optical theorem preserve the standard
null-test logic for T-odd forward neutron transmission? The answer is yes, provided that ``time
reversal'' is implemented in the full prepared mode space. Let $\Theta$ be the antiunitary time-reversal operator. \TRI{} implies
\begin{equation}
\Theta S\Theta^{-1}=S^{-1}=S^{\dagger},
\label{eq:TRI_S}
\end{equation}
and therefore, using \cref{eq:S_def},
\begin{equation}
\Theta T\Theta^{-1}=T^{\dagger}.
\label{eq:TRI_T}
\end{equation}
Take an arbitrary prepared incident state $\ket{\psi_{\instate}}$ and define its time-reversed
partner by
\begin{equation}
\ket{\psi^T_{\instate}}\equiv\Theta\ket{\psi_{\instate}}.
\label{eq:psiT}
\end{equation}
Starting from \cref{eq:hilbert_ot},
\begin{equation}
2\Im\,\mel{\psi_{\instate}}{T}{\psi_{\instate}}
=\sum_f\left|{}_{\outstate}\mel{f}{T}{\psi_{\instate}}\right|^2,
\label{eq:ot_repeat}
\end{equation}
we note that antiunitarity preserves absolute squares, so that
\begin{equation}
\left|{}_{\outstate}\mel{f}{T}{\psi_{\instate}}\right|^2
=\left|{}_{\outstate}\mel{\Theta f}{T}{\Theta\psi_{\instate}}\right|^2,
\label{eq:antiunitary_sq}
\end{equation}
where \cref{eq:TRI_T} has been used. Relabeling the complete out-basis therefore gives
\begin{equation}
\sum_f\left|{}_{\outstate}\mel{f}{T}{\psi_{\instate}}\right|^2
=\sum_f\left|{}_{\outstate}\mel{f}{T}{\psi^T_{\instate}}\right|^2,
\label{eq:inclusive_equal}
\end{equation}
and hence
\begin{equation}
2\Im\,\mel{\psi_{\instate}}{T}{\psi_{\instate}}
=2\Im\,\mel{\psi^T_{\instate}}{T}{\psi^T_{\instate}}.
\label{eq:TRI_forward_equal}
\end{equation}
With the same flux convention for the two compared preparations,
\begin{equation}
\sigma_{\tot}[\psi_{\instate}]=\sigma_{\tot}[\psi^T_{\instate}].
\label{eq:sigma_equal_TRI}
\end{equation}
Equation~\eqref{eq:sigma_equal_TRI} is the generalized forward-transmission null theorem. It is the
same physical statement that underlies the classic Ryndin result and its modern neutron-transmission
formulation, but now expressed directly in the full mode space appropriate to structured beams
\cite{BilenkiiLapidusRyndin1965,BilenkiiLapidusRyndin1969,BowmanGudkov2014}.

\subsection{What time reversal does to neutron OAM modes}

For a massive neutron,
\begin{equation}
\mathbf k\to-\mathbf k,
\qquad
\mathbf L\to-\mathbf L,
\qquad
\mathbf S\to-\mathbf S.
\label{eq:TR_actions}
\end{equation}
Accordingly, for an OAM mode defined about the beam axis the winding quantum number changes as
$m\to -m$ up to phase conventions. A structured-beam null test must therefore compare two prepared
initial states related by time reversal in their full mode content, including the momentum
distribution, spin, OAM content, and any experimentally relevant acceptance definition. Merely
``flipping the spin'' is not enough if the mode preparation itself is not also the correct time
reverse.

\subsection{Operational statement for transmission experiments}

The theorem does \emph{not} say that any pair of nominally similar apparatus settings must produce
identical forward loss. It says that if the two compared preparations are genuine time reverses in
the full mode space, then \TRI{} forbids a false T-odd signal. This distinction matters because a
target can scatter amplitude out of the originally prepared mode into other modes, including
OAM-carrying ones. Schematically,
\begin{equation}
S\ket{\psi_1}=\alpha\ket{\psi_1}+\sum_{\nu\neq 1}\beta_{\nu}\ket{\psi_{\nu}},
\qquad
\sum_{\nu}|\beta_{\nu}|^2=1-|\alpha|^2.
\label{eq:mode_leakage}
\end{equation}
The strict time-reversed counterpart of setting 1 is therefore not simply ``flip a field'' or ``flip
a spin,'' but rather the preparation of the time reverse of the full outgoing state,
\begin{equation}
\ket{\psi_{\instate}^{(\mathrm{TR})}}=\Theta S\ket{\psi_1}.
\label{eq:true_TR_state}
\end{equation}
If an experiment compares two settings that differ in spin or field configuration but do not control
the full mode content, then the compared states need not be true time reverses. A nonzero
transmission difference can then appear even when \TRI{} holds. That is not a loophole in the
Ryndin theorem; it is a mode-preparation systematic.

\section{Mode mismatch and OAM leakage as the practical systematic}
\label{sec:mismatch}

The previous section identifies the formal requirement for an exact null test. The natural next
question is how large a false signal can arise when the implemented comparison deviates slightly from
the true time-reversed pairing. Let the ideal pair satisfy
\begin{equation}
\ket{\psi_2^{(\mathrm{ideal})}}=\Theta\ket{\psi_1},
\label{eq:ideal_pair}
\end{equation}
but suppose that the actually prepared second state is
\begin{equation}
\ket{\psi_2}=\Theta\ket{\psi_1}+\ket{\delta\psi},
\qquad
\norm{\delta\psi}\equiv\epsilon\ll 1.
\label{eq:delta_psi_def}
\end{equation}
Define the inclusive forward-loss functional
\begin{equation}
L[\psi]\equiv 2\Im\,\mel{\psi}{T}{\psi},
\label{eq:L_def}
\end{equation}
so that, after the flux convention is inserted, $L$ is proportional to the total transmission loss.
The null observable is
\begin{equation}
\Delta L\equiv L[\psi_1]-L[\psi_2].
\label{eq:deltaL_def}
\end{equation}
Under exact \TRI{} and exact mode reversal, $\Delta L=0$. Expanding to first order in
$\ket{\delta\psi}$ gives
\begin{align}
L[\psi_2]
&=2\Im\mel{\Theta\psi_1+\delta\psi}{T}{\Theta\psi_1+\delta\psi}
\notag\\
&=2\Im\mel{\Theta\psi_1}{T}{\Theta\psi_1}
+4\Im\mel{\delta\psi}{T}{\Theta\psi_1}+\mathcal O(\epsilon^2).
\label{eq:L_expand}
\end{align}
Using the exact \TRI{} equality for the ideal pair,
\begin{equation}
\Delta L=-4\Im\mel{\delta\psi}{T}{\Theta\psi_1}+\mathcal O(\epsilon^2).
\label{eq:deltaL_first_order}
\end{equation}
A conservative norm bound follows immediately from Cauchy--Schwarz,
\begin{equation}
|\Delta L|\le 4\,\norm{\delta\psi}\,\norm{T\Theta\psi_1}+\mathcal O(\epsilon^2)
\le 4\epsilon\norm{T}+\mathcal O(\epsilon^2).
\label{eq:deltaL_bound}
\end{equation}
Thus false null-test violations scale linearly with the mode-mismatch amplitude unless an additional
symmetry suppresses the linear term.

The same point can be expressed in the moment language of \cref{eq:feff_operator_final}. If the
intended angular weight is $w(\Omega)$ but the actual weight is $w(\Omega)+\delta w(\Omega)$ with
\begin{equation}
\int\dd\Omega\,\delta w(\Omega)=0,
\end{equation}
then to leading order
\begin{equation}
\delta f_{\eff}=\int\dd\Omega\,\delta w(\Omega)\,f\bigl(\khat(\Omega)\bigr).
\label{eq:deltafeff}
\end{equation}
The induced false shift in any extracted coefficient is therefore controlled by the corresponding
uncertainty in beam moments such as $\delta\avg{\khat}$ and
$\delta\avg{\khat\khat^{\mathsf T}}$. This is the natural way to propagate structured-beam
systematics in an actual transmission analysis. 

Here we give an example to show that the change in the neutron spin is compensated by the generation of neutron orbital angular momentum and use it to provide a rough estimate of the size of the effect. The neutron-nucleus weak interaction Hamiltonian in the optical limit is given by the anti-commutator between the weak potential, $F(r)$, and the neutron helicity operator $\vec{\sigma}\cdot\vec{p}$

$$
H=\{F(r),\vec{\sigma}\cdot\vec{p}\}=F(r)\vec{\sigma}\cdot\vec{p}-iF_r(r)\vec{\sigma}\cdot\hat{r}$$

where the subscript $r$ indicates the first derivative to r. The neutron helicity operator causes the parity odd spin rotation, while the $\sigma\cdot\hat{r}$ term is the source of parity odd spin dichroism~\cite{Haseyama:2001mg}. The spin rotation term conserves total angular momentum. Writing $\vec{\sigma}\cdot\vec{p}=k_z\sigma_z+k_r\sigma_+\ell_-+k_r\sigma_-\ell_+$ reveals that a mode entanglement between spin and OAM takes place in this interaction, and with slightly more involved calculation it can be shown that even longitudinal spin-orbit states can be generated when the spin is parallel to the momentum vector. These states have a small amplitude equal to $\theta \frac{k_r}{k_z}$, with $\theta$ the parity odd spin rotation in the case where the initial spin is orthogonal to the momentum vector and $\frac{k_r}{k_z}$ is equal to the beam divergence. In practice therefore the potential systematic error that might be induced by this OAM generation process is the same size as those which are already present from the finite beam divergence which must be suppressed in any case, and therefore in practice it will not generate a new systematic error type. 

\section{Discussion and outlook}
\label{sec:discussion}

The central message of this paper is conceptually simple. The optical theorem formula $\sigma_{\tot}=(4\pi/k)\Im f(0)$ is the plane-wave limit of a
more general same-mode unitarity statement. For a structured beam of a massive nonrelativistic spin-$1/2$ particle, the correct forward quantity is the amplitude to return to the same prepared incoming mode, or its appropriate mode average in bulk transmission.

For neutron optics this result matters because forward transmission is precisely where the optical theorem and the Ryndin null-test logic are most tightly connected. The physical definition of the total cross section remains unchanged: it is the inclusive rate of removal from the prepared mode divided by the incident flux. What changes is only the forward amplitude that appears in the theorem. For coherent structured states it is $F[a]$; for bulk transmission it is the mode-effective operator $f_{\eff}[w]$. The forward operator keeps the same spin structures, but explicit dependence on
$\khat$ is replaced by beam moments.

The second main result is that the forward-transmission null theorem survives intact as long as the neutron state preparations that are compared are true time reverses in the full mode space: then \TRI{} forbids a false T-odd forward signal even for structured beams. Accordingly, OAM beams do not by itself invalidate the null-test theorem. The practical issue of mode control is however relevant for the experimental realization: OAM leakage, imperfect reversal of the prepared angular distribution, and acceptance mismatch can all mimic a null-test signal if they cause the compared preparations to cease being exact time reverses of each other. Fortunately this  systematic effect in a realistic apparatus is folded into the effects from finite beam divergence and does not cause a new systematic.

Although the presentation here has emphasized neutrons, the formal result is more general and applies to any massive nonrelativistic spin-$1/2$ particle prepared in a structured incoming mode. The application to neutron optics, which routinely uses forward scattering as an operational language, is straightforward. Neutron OAM is now in the process of becoming experimentally accessible and can provide a physical setting in which the generalized optical theorem developed here becomes experimentally relevant. Various neutron-nucleus interactions, including the parity violation example above, can themselves generate neutron OAM, and calculations of OAM generation on neutron--nucleus resonances are in progress. 

\acknowledgments

The author thanks W. M. Snow and A. Afanasev for discussions on the subject of this paper during the workshop \lq\lq Twisted Light in Nanophotonic and Quantum Systems\rq\rq, held at The George Washington University, Washington DC, December 17-19, 2025. The author acknowledges support from US National Science Foundation grant PHY-2209481, the US Department of Energy grant DE-SC0023695, and the Indiana University Center for Spacetime Symmetries.  

\appendix

\section{Diagonalization of the structured-beam forward functional}
\label{app:diag}

For completeness we record the algebra leading from \cref{eq:F_def} to \cref{eq:F_diag}. Substitute
\begin{equation}
a(\Omega)=\sum_{\ell m}a_{\ell m}Y_{\ell m}(\Omega)
\end{equation}
and the kernel expansion implied by \cref{eq:f_legendre,eq:addition_theorem},
\begin{equation}
K(\Omega,\Omega')=4\pi\sum_{\ell m}f_{\ell}Y_{\ell m}(\Omega)Y_{\ell m}^*(\Omega').
\end{equation}
Then
\begin{align}
F[a]
&=\int\dd\Omega\int\dd\Omega'\sum_{\ell m}a_{\ell m}^*Y_{\ell m}^*(\Omega)
\,4\pi\sum_{\ell' m'}f_{\ell'}Y_{\ell' m'}(\Omega)Y_{\ell' m'}^*(\Omega')
\sum_{\ell''m''}a_{\ell''m''}Y_{\ell''m''}(\Omega')
\notag\\
&=4\pi\sum_{\ell m}\sum_{\ell' m'}\sum_{\ell''m''}
 a_{\ell m}^*f_{\ell'}a_{\ell''m''}
\left(\int\dd\Omega\,Y_{\ell m}^*Y_{\ell' m'}\right)
\left(\int\dd\Omega'\,Y_{\ell' m'}^*Y_{\ell''m''}\right)
\notag\\
&=4\pi\sum_{\ell m}f_{\ell}|a_{\ell m}|^2,
\end{align}
which is \cref{eq:F_diag}.

\section{Axial-cone moments}
\label{app:cone}

For the axial cone of \cref{eq:cone_weight}, the unit vector is
\begin{equation}
\khat=(\sin\theta_0\cos\phi,\sin\theta_0\sin\phi,\cos\theta_0).
\end{equation}
Averaging over azimuth gives
\begin{align}
\avg{k_x}&=0,
&\avg{k_y}&=0,
&\avg{k_z}&=\cos\theta_0,
\\
\avg{k_x^2}&=\frac{\sin^2\theta_0}{2},
&\avg{k_y^2}&=\frac{\sin^2\theta_0}{2},
&\avg{k_z^2}&=\cos^2\theta_0,
\end{align}
while all mixed moments vanish by symmetry. These relations imply
\cref{eq:first_cone_moment,eq:second_cone_moment}. Equation~\eqref{eq:F_cone} then follows by
direct substitution into \cref{eq:moment_tensor_identity}.

\section{Channel-space form for noncentral spin-dependent interactions}
\label{app:noncentral}

The T-odd structures relevant to neutron transmission are not generated by a purely central scalar
interaction, so it is useful to record the corresponding channel-space form of the generalized
optical theorem. Let the asymptotic channel label $\alpha$ denote whatever discrete quantum numbers
are needed to specify the coupled problem, for example orbital partial wave, spin projection,
total angular momentum, or any other convenient basis. A general incoming structured state may then
be written as
\begin{equation}
\ket{\Psi_{\instate}}=\sum_{\alpha}\int\dd\Omega\,a_{\alpha}(\Omega)\,
\ket{\Omega,\alpha}_{\instate},
\qquad
\sum_{\alpha}\int\dd\Omega\,|a_{\alpha}(\Omega)|^2=1.
\label{eq:channel_state}
\end{equation}
The scattering kernel becomes an operator in channel space,
\begin{equation}
K_{\alpha\beta}(\Omega,\Omega')\equiv {}_{\outstate}\mel{\Omega,\alpha}{T}{\Omega',\beta}_{\instate},
\label{eq:channel_kernel}
\end{equation}
and the same-mode forward functional is correspondingly
\begin{equation}
F[a]=\sum_{\alpha\beta}\int\dd\Omega\int\dd\Omega'\,
 a_{\alpha}^*(\Omega)K_{\alpha\beta}(\Omega,\Omega')a_{\beta}(\Omega').
\label{eq:F_channel}
\end{equation}
The Hilbert-space identity \cref{eq:hilbert_ot} then implies, after the same flux convention is
chosen as in the main text,
\begin{equation}
\sigma_{\tot}[a]=\frac{4\pi}{k}\Im F[a],
\label{eq:channel_ot}
\end{equation}
with the understanding that $F[a]$ is now built from the full operator-valued kernel.

In a basis that diagonalizes the coupled-channel $S$ matrix, one may write
\begin{equation}
F[a]=\sum_{n}\lambda_n\,|c_n|^2,
\label{eq:channel_diag}
\end{equation}
where $n$ labels the eigenchannels and $c_n$ are the overlaps of the prepared incident mode with the
corresponding eigenvectors. This is the natural noncentral analogue of \cref{eq:F_diag}. The same
logic therefore survives unchanged: the optical theorem remains a same-mode unitarity statement, but
the relevant forward object is a quadratic form in the full coupled channel space rather than a
single scalar partial-wave sum.

For forward neutron transmission one often packages the channel dependence into an effective forward
spin operator such as \cref{eq:forward_operator_plane}. The mode-averaged operator
\cref{eq:feff_operator_final} should be viewed as the low-order neutron-optics representation of this
more general channel-space statement.

\bibliography{refs}

\end{document}